\documentclass[%
 preprint,
 nofootinbib,
 amsmath,amssymb,
 aps,
 a4paper
]{revtex4-1}

\usepackage{graphicx}% Include figure files
\usepackage{dcolumn}% Align table columns on decimal point
\usepackage{bm}% bold math
\usepackage{slashed}
\usepackage{color}
\begin{document}

\preprint{UT-19-23}

\title{Aspects of Nonlinear Effect on Black Hole Superradiance}

\author{Hajime Fukuda}
\email{hfukuda@hep-th.phys.s.u-tokyo.ac.jp}
 \affiliation{Department of Physics, Faculty of Science,
The University of Tokyo,  Bunkyo-ku, Tokyo 113-0033, Japan}
\affiliation{Theoretical Physics Group, Lawrence Berkeley National Laboratory,
 CA 94720, USA}
\affiliation{Berkeley Center for Theoretical Physics, Department of Physics,\\
 University of California, Berkeley, CA 94720, USA}

\author{Kazunori Nakayama}%
 \email{kazunori@hep-th.phys.s.u-tokyo.ac.jp}
\affiliation{Department of Physics, Faculty of Science,
The University of Tokyo,  Bunkyo-ku, Tokyo 113-0033, Japan}%
\affiliation{Kavli IPMU (WPI), The University of Tokyo, Kashiwa, Chiba 277-8583, Japan}%

\date{\today}% It is always \today, today,

\begin{abstract}

Under some conditions, light boson fields grow exponentially around a rotating black hole, called the superradiance instability. We discuss effects of nonlinear interactions of the boson on the instability. In particular, we focus on the effect of the particle production and show that the growth of the boson cloud may be saturated much before the black hole spin is extracted by the boson cloud, while the nonlinear interactions also induce the boson emission. For application, we revisit the superradiant instability of the standard model photon, axion and hidden photon. 

\end{abstract}

\maketitle

%\tableofcontents

%%%%%%%%%%%%%%%%%%%%%%%%%%%%%%%%%%%%%%
\section{Introduction}
%%%%%%%%%%%%%%%%%%%%%%%%%%%%%%%%%%%%%%

There may exist light scalar fields in theories beyond the standard model~\cite{Arvanitaki:2009fg} and many ideas are proposed to find signatures of such light particles including terrestrial experiments and astrophysical observations. 
One of the ideas is to see the effects of light particles on the black hole physics. As we briefly review below, light bosons around a rotating black hole (or Kerr black hole) may experience a so-called superradiant instability and the boson cloud may be formed. 
It can significantly affect the evolution of the black hole through the extraction of its mass and spin by the boson cloud, which can be severely constrained by observations. 
Theoretical and phenomenological aspects of black hole superradiance are found in Refs.~\cite{Press:1972zz,Bekenstein:1973mi,Damour:1976kh,Zouros:1979iw,Detweiler:1980uk,Furuhashi:2004jk,Dolan:2007mj,Rosa:2009ei,Arvanitaki:2010sy,Rosa:2011my,Yoshino:2012kn,Pani:2012vp,Pani:2012bp,Dolan:2012yt,Pani:2013hpa,Yoshino:2013ofa,Arvanitaki:2014wva,Brito:2014wla,Brito:2015oca,Endlich:2016jgc,East:2017ovw,Baryakhtar:2017ngi,Cardoso:2018tly,Baumann:2018vus,Dolan:2018dqv,East:2018glu,Baumann:2019eav}.

A massive scalar field $\phi$ with its mass $\mu$ around a black hole satisfies the Klein-Gordon equation
\begin{align}
	(\square - \mu^2) \phi=0.
\end{align}
Under the Kerr metric, the solution to this equation of the form $\phi\propto e^{-i\omega t+ im\varphi}$ is found, where $m$ is a quantum number corresponds to angular momentum around the rotating axis and $\varphi$ is the azimuthal angle. The frequency $\omega=\omega_R+ i\omega_I$ is given by
\begin{align}
	&\omega_R \simeq \mu\left(1-\frac{\alpha^2}{2(n+\ell+1)^2} \right),\\
	&\omega_I \simeq \frac{1}{\gamma_{n\ell m} GM_{\rm BH}} \left(\tilde a m-2\mu r_+ \right)(GM_{\rm BH} \mu)^{4\ell+5},
	\label{omegaI}
\end{align}
for $GM_{\rm BH}\mu\ll 1$, where $G$ is the Newton constant, $M_{\rm BH}$ is the black hole mass, $\tilde a \equiv a/(GM_{\rm BH})$ is the dimensionless spin parameter in the range $0\leq \tilde a \leq 1$ with $a$ being the parameter appearing in the Kerr metric, which is related to the black hole angular momentum $J_{\rm BH}$ through $a=J_{\rm BH}/M_{\rm BH}$, $r_+= GM_{\rm BH} + \sqrt{(GM_{\rm BH})^2-a^2}$ represents the event horizon and $\gamma_{n\ell m}$ is a numerical constant for the principal quantum number $n$ and orbital angular momentum quantum number $\ell$. We have also defined a dimensionless quantity $\alpha \equiv GM_{\rm BH}\mu$ for convenience.\footnote{
	The factor $(\tilde a m-2\mu r_+)$ in Eq.~(\ref{omegaI}) can be rewritten as $2r_+(m\Omega_{\rm H} - \mu)$ by using the  angular velocity of black hole event horizon, $\Omega_{\rm H} =a/(r_+^2+a^2)$.
}
It is seen that if $a$ is larger than the critical value $a_{\rm crit} = 2r_+ GM_{\rm BH}\mu/m + \mathcal{O}(\alpha^2)$, $\omega_I$ is positive and the instability happens. This is called the superradiant instability. 
The growth rate is maximized for the mode $(n,\ell,\mu)=(0,1,1)$, $\tilde a \simeq 1$ and $\alpha \simeq 0.42$: in this case we have $\omega_I \sim 10^{-7} \mu$. For this reason, the typical time scale of the superradiant instability\footnote{In reality, it roughly takes $\ln(\phi_\text{max} / \mu) \sim 100$ times more for the boson cloud to be formed from the vacuum fluctuation and extract the angular momentum, where $\phi_\text{max}$ is the maximal amplitude of the cloud. 
%We do not show this factor in the equation, but include it in the figures in the rest of this paper.
Thus we define the superradiance time scale as $\tau_{\rm SR} =\ln(\phi_\text{max} / \mu)\,\omega_I^{-1}$ and use it in figures in the rest of this paper. 
}\label{footnote:SR} is taken to be $\omega_I^{-1}$. Note that $\omega_I$ is a very steep function of the combination $GM_{\rm BH} \mu$ and hence the instability soon becomes inefficient for smaller $GM_{\rm BH} \mu$.  Numerically we have
\begin{align}
	GM_{\rm BH} \mu \simeq \left(\frac{\mu}{1.3\times 10^{-10}\,{\rm eV}}\right)\left(\frac{M_{\rm BH}}{M_\odot}\right).
\end{align}
For astrophysical black holes $M_{\odot}\lesssim M_{\rm BH} \lesssim 10^{10}M_\odot$, for example, the target scalar mass is $10^{-11}\,{\rm eV} \gtrsim \mu \gtrsim 10^{-21}\,{\rm eV}$.

A similar superradiant instability happens also for vector bosons. In this cases the frequency is calculated as~\cite{Rosa:2011my,Pani:2012vp,Pani:2012bp,Brito:2015oca,Baryakhtar:2017ngi}
\begin{align}
	&\omega_R \simeq \mu\left(1-\frac{\alpha^2}{2(n+\ell+1)^2}\right),\\
	&\omega_I \simeq \frac{\gamma_{j\ell}}{GM_{\rm BH}} \left( \tilde a m-2\mu r_+ \right)(GM_{\rm BH} \mu)^{2 j + 2\ell+5},
	\label{omegaI_vec}
\end{align}
for $GM_{\rm BH} \mu\ll 1$, where $j$ is the total angular momentum and  $\gamma_{j\ell}$ denotes a numerical constant. 
It reaches a maximum growing rate $\omega_I \sim 10^{-3} \mu$ for $(\ell,j)=(0,1)$, $\tilde a\simeq 1$ and $\alpha \sim 0.5$~\cite{Baryakhtar:2017ngi,Dolan:2018dqv}.

The above analysis shows that if there exists a light scalar or vector boson, it experiences a superradiant instability around the near-extremal Kerr black hole and the boson cloud is formed. The instability continues until a significant fraction of black hole mass or spin is extracted by the boson cloud and $a$ becomes smaller than $a_\text{crit}$.
Thus the measurement of the black hole spin can constrain the existence of such a light scalar or vector boson~\cite{Brito:2014wla,Brito:2015oca}.

So far, it has been assumed that the boson is a free field, i.e., it has only a gravitational interaction. However, it is often the case that a boson has interactions with other fields.
A representative model of a light scalar is the axion-like particle, whose potential often appears from some non-perturbative effect and looks like $V(\phi) \sim \mu^2f^2(1-\cos(\phi/f))$ with axion decay constant $f$. In this case, the axion has nonlinear self interactions.
The vector boson also usually has gauge interactions with some other fields including charged matter or the Higgs boson.
In general, if we neglect any nonlinear interaction, the total mass of the cloud when a substantial fraction of the black hole spin is extracted is $M_{\rm cloud} \sim \tilde a \alpha M_{\rm BH} / m$, which implies that the typical field amplitude in the boson cloud is
\begin{align}
\left<\phi^2\right> \sim \frac{\tilde a \alpha M_{\rm BH}}{m \mathcal V \mu^2} \sim 8 \tilde a \alpha^5 M_{\rm Pl}^2,
\end{align}
where $M_{\rm Pl}$ denotes the reduced Planck scale and $\mathcal V \sim \pi (\alpha \mu)^{-3}$ is the effective volume of the boson cloud. This shows that the field amplitude is not very far from the Planck scale, Thus, it is reasonable to expect that nonlinear effects plays important roles before a significant fraction of the black hole spin is extracted.
These nonlinear interactions can drastically modify phenomenological consequences of the superradiance.
It would be a very complicated task to precisely solve the dynamics including the nonlinearity in general, but we can still obtain a reasonable estimate for the effect of the nonlinearity.
In particular, we focus on the effect of the particle production on the superradiant instability and its phenomenological implications.

The rest of the paper is organized as follows.
In Sec.~\ref{sec:nonlinear}, the rough picture of nonlinear effects on the superradiance is explained. 
In Sec.~\ref{sec:app} we consider some phenomenological implications of such nonlinear effects. The Schwinger pair production of the standard model photon around the primordial black hole and the particle creation by nonlinear self-interactions of axion-like particle, hidden photon and generically interacting scalar is discussed.
Sec.~\ref{sec:dis} is devoted to conclusions and discussion.

%%%%%%%%%%%%%%%%%%%%%%%%%%%%%%%%%%%%%%
\section{Nonlinear effects on superradiant instability}   \label{sec:nonlinear}
%%%%%%%%%%%%%%%%%%%%%%%%%%%%%%%%%%%%%%

%%%%%%%%%%%%%%%%%%%%%%%%%%%%%%%%%%%%%%
\subsection{Rough sketch}
%%%%%%%%%%%%%%%%%%%%%%%%%%%%%%%%%%%%%%

As explained above, if there is a bosonic particle, $\phi$, and the mass is the same order as the horizon radius of a Kerr black hole, the rotational energy of the Kerr black hole is efficiently extracted by $\phi$ by the superradiant instability. Obtaining the rotational energy, the $\phi$ cloud emerges around the black hole. The amplitude of the cloud exponentially grows up. 
As a result, a substantial fraction of the black hole energy and angular momentum are transferred to the boson cloud.
%In other words, the exponentially dense $\phi$ cloud is necessary for the efficient rotation energy extraction.

However, as the boson cloud grows, the nonlinear interactions become important and may affect the superradiant exponential growth. The following  points need to be taken into consideration in order to figure out how the nonlinearity affects the superradiance process.
\begin{enumerate}
    % \item Nonlinear effective potential arises from interactions and it affects the efficiency of superradiance.
    \item Self-interactions may produce $\phi$ particles with higher momenta.
    \item Other particles interacting with $\phi$ may be produced.
    \item The bound state spectrum between $\phi$ and the black hole may be changed.
\end{enumerate}
In this paper, we mainly focus on the first two points. We may take the spectrum distortion into the account as the change of the effective mass, but in the following models we discuss, it turns out that the particle creation process becomes already effective before the effective mass significantly changes.
% In this paper, we focus on the first three points.
% The change of the spectrum can be described in terms of the change of the effective mass in the lowest order approximation in the coupling, but later we will see that once the amplitude becomes so large that the nonlinearity becomes important and the effective mass is changed by $\mathcal O(1)$, the particle creation process becomes automatically important in the most parameter region of interest.

The particle creation process leads to the energy leakage from the $\phi$ cloud surrounding the black hole.
If the energy leakage rate by the particle creation becomes equivalent to the energy extraction rate by superradiance at a given amplitude, $\phi_{\rm NL}$, above which the leakage rate is larger, the extracted energy can be considered to be dominantly converted into the created particle. Then, the cloud does not grow up and thus the amplitude of $\phi$ cannot be larger than $\phi_{\rm NL}$. 
%The reason why the superradiance process is effective to extract the black hole angular momentum is that the extraction rate grows exponentially. 
Once the exponential growth of the amplitude stops at $\phi_{\rm NL}$, so does the energy/angular momentum extraction rate by the superradiant instability. In such cases, the energy/angular momentum loss rate of the black hole becomes saturated and constant in time. Compared with the free boson superradiant instability, where the loss rate exponentially grows up, the typical time scale needed for a substantial energy/angular momentum extraction becomes significantly lengthen.

Hence, in order to calculate the black hole spinning down time correctly, we need to estimate the particle creation rate in the boson cloud. In the next section, we derive the particle production rate for several models with nonlinear interactions.
Before going into concrete setups, we below summarize some general aspects of the black hole evolution taking account of nonlinear effects. 
%We assume that the wavelength of the created particle is much larger than the mass of $\phi$ and thus the background Kerr metric may be ignored.
%We ignore the rotational motion of the $\phi$ cloud and simply assume that the $\phi$ background behaves as $\sim e^{-i\omega t}$, where $\text{Im}\,\omega > 0$ because of the superradiance instability.

%%%%%%%%%%%%%%%%%%%%%%%%%%%%%%%%%%%%%%
\subsection{Time evolution of black hole}
%%%%%%%%%%%%%%%%%%%%%%%%%%%%%%%%%%%%%%

Let us consider the system of a rotating black hole and the boson cloud surrounding it, which is formed by the superradiant instability.
The mass and angular momentum of the rotating black hole are denoted by $M_{\rm BH}$ and $J_{\rm BH}$ and those of cloud consisting of light scalar/vector boson are denoted by $M_{\rm cloud}$ and $J_{\rm cloud}$, respectively. The angular momentum of the could is given by $J_{\rm cloud} = (m/\mu) M_{\rm cloud}$. 
The time evolution of the cloud is described by
\begin{align}
	&\dot M_{\rm cloud} = 2 \omega_I M_{\rm cloud}-\dot M_{\rm NL}\\
	&\dot J_{\rm cloud} = 2 \omega_I J_{\rm cloud}-\dot J_{\rm NL},
\end{align}
where $\dot M_{\rm NL}$ and $\dot J_{\rm NL}$ represent the energy and angular momentum extraction rate due to nonlinear effects, respectively.\footnote{
	The effect of gravitational wave emission is safely neglected in the discussion as far as the nonlinear effect becomes important much before the field amplitude grows to the Planck scale.
}
Similarly, the time evolution of the black hole is described by
\begin{align}
	&\dot M_{\rm BH} = -2 \omega_I M_{\rm cloud}+\dot M_{\rm acc}\\
	&\dot J_{\rm BH} = -2 \omega_I J_{\rm cloud}+\dot J_{\rm acc},
\end{align}
where $\dot M_{\rm acc}$ and $\dot J_{\rm acc}$ represent the accretion from the surrounding matter, respectively.
The total mass and angular momentum of the black hole-cloud system are: $M_{\rm tot} = M_{\rm BH} + M_{\rm cloud}$ and $J_{\rm tot} = J_{\rm BH} + J_{\rm cloud}$. Their time evolution are thus governed by
\begin{align}
	&\dot M_{\rm tot} = \dot M_{\rm acc} - \dot M_{\rm NL},\\
	&\dot J_{\rm tot} = \dot J_{\rm acc} - \dot J_{\rm NL}.
\end{align}

The accretion rate depends on the environment around the black hole.
However, there is an upper bound, the Eddington limit, at which the gravitational infall into the black hole and the radiation pressure from the falling matter is balanced. If this bound is saturated, the typical accretion time scale is 
\begin{align}
	\tau_{\rm acc} \simeq \frac{M_{\rm BH}}{\dot M_{\rm acc}} = \frac{\epsilon \sigma_{\rm T}}{4\pi G m_p} \simeq 1.4\epsilon \times 10^{15}\,{\rm sec},
	 \label{t_acc}
\end{align}
which is independent of the black hole mass, where $m_p$ is the proton mass, $\sigma_{\rm T}$ is the Thomson scattering cross section for the electron, $\epsilon$ is the radiative efficiency. For the near-extremal Kerr black hole, $\epsilon \sim 0.3$\,\cite{Thorne:1974ve}. % and we have substituted $\dot M_{\rm acc}=4\pi GM_{\rm BH} m_p/\epsilon \sigma_{\rm T}$, which is the Eddington luminosity. 
If the Eddington limit is not saturated, the accretion time scale can be much longer.

First, let us suppose that there are no nonlinear effects and initially the superradiance is inefficient: $GM_{\rm BH} \mu\ll 1$. The total mass and angular momentum gradually increase due to the accretion and the superradiant instability becomes effective around the epoch $GM_{\rm BH} \mu\sim \mathcal O(1)$ for the lowest-excited mode with $m = 1$. The typical time scale of the superradiant instability, $\omega_I^{-1} \sim 10^{3\mbox-7} \mu^{-1}$, is very short compared with the accretion time scale. Thus, the boson cloud forms rapidly, converting a significant fraction of black hole mass and spin into the cloud. Since
\begin{align}
	\frac{\dot a}{a} = \frac{\dot J_{\rm BH}}{J_{\rm BH}} - \frac{\dot M_{\rm BH}}{M_{\rm BH}} = -\frac{2\omega_I M_{\rm cloud}}{M_{\rm BH}}\left( \frac{m}{\tilde a\alpha}-1 \right),
\end{align}
and the factor in the parenthesis is positive, the spin parameter $a$ is decreasing through this process. The superradiant instability stops when the spin $a$ becomes smaller than the critical value, $a_{\rm crit}$~\cite{Bekenstein:1973mi,Brito:2014wla,Brito:2015oca}.
Typically, a black hole with dimensionless spin parameter $\tilde a\simeq 1$ loses its spin by $\mathcal O(1)$ fraction: $\Delta \tilde a \sim \tilde a$. The mass of the cloud at this stage is $M_{\rm cloud} \sim \mu J_{\rm cloud} \sim \mu \Delta J_{\rm BH} \sim (\Delta \tilde a)\alpha M_{\rm BH}$.
After that, the mass and spin of black hole again increase due to the accretion and the superradiant instability becomes inefficient.
Thus, there would appear forbidden region on the black hole Regge plane ($M_{\rm BH}$ v.s. $\tilde a$), which can be compared with observations. It can give constraints on light scalar and vector bosons with mass range of $10^{-20}\,{\rm eV}$--$10^{-11}\,{\rm eV}$~\cite{Brito:2014wla,Brito:2015oca}.

The story drastically changes if one takes account of nonlinear effects. As discussed in the previous section, the growth of the boson cloud may first stop much before it extracts the significant fraction of the black hole mass and spin. The decrease of the black hole mass and spin due to the superradiance at this point can be completely negligible, i.e., they are saturated at $M_\text{cloud}\ll M_{\rm BH}$ and $J_\text{cloud} \ll J_{\rm BH}$, so that the observation may not directly constrain the existence of such light bosons. However, here the second effect may take an important role: the nonlinear interactions extract the boson cloud energy through the production of other particles or emission of high frequency modes, represented by $\dot M_{\rm NL}$ and $\dot J_{\rm NL}$. 
Therefore, if the accretion rate is smaller than the extraction rate, the system gradually loses mass and angular momentum. 
Although it may be much less efficient than the case without nonlinear interaction, it is constrained from observations in principle. 

In the next section we show some concrete examples in which nonlinear effects play essential roles to discuss the phenomenological consequences of the superradiant instability.

%%%%%%%%%%%%%%%%%%%%%%%%%%%%%%%%%%%%%%
\section{Examples}   \label{sec:app}
%%%%%%%%%%%%%%%%%%%%%%%%%%%%%%%%%%%%%%

%%%%%%%%%%%%%%%%%%%%%%%%%%%%%%%%%%%%%%
\subsection{Standard Model photon around primordial black hole}  \label{sec:SM}
%%%%%%%%%%%%%%%%%%%%%%%%%%%%%%%%%%%%%%

Here, we revisit the constraint on the primordial black hole (PBH) abundance from the superradiance of the standard model electromagnetic photon\,\cite{Pani:2013hpa}.
Since the photon obtains an effective mass in the ionized plasma, called the plasma frequency $\omega_p$, the superradiant instability may happen if there is a PBH with its mass satisfying $GM_{\rm BH}\omega_p = \alpha$ $(\simeq 0.5)$.\footnote{
	The initial spin of the PBH $\tilde a$ may be typically percent level~\cite{Chiba:2017rvs,Mirbabayi:2019uph,DeLuca:2019buf} if the PBH is formed at the radiation-dominated era. In this case, the effect of superradiance around PBHs may be extremely small since $a> a_{\rm crit}$ requires small $GM_{\rm BH}\mu$ that greatly suppresses $\omega_I$. If the PBH is formed at the matter-dominated era, on the other hand, the initial spin can be large~\cite{Harada:2017fjm}. Below we assume that $\tilde a$ is at least $\mathcal O(0.1)$.
} The plasma frequency is given by
\begin{align}
	\omega_p = \left( \frac{4\pi \alpha_e n_e}{m_e} \right)^{1/2} \simeq 2\times 10^{-8}\,{\rm eV}\left( \frac{1+z}{10^4} \right)^{3/2} X_e^{1/2},
\end{align}
where $X_e$ is the ionized fraction of the hydrogen. For a given PBH mass $M_{\rm BH}$, the condition $GM_{\rm BH}\omega_p = \alpha$ is satisfied at the redshift
\begin{align}
	1+z_{M} \sim 4\times 10^2 \left(\frac{M_\odot}{M_{\rm BH}}\right)^{2/3}X_e^{-1/3} \alpha^{2/3}.  \label{z_M}
\end{align}
Thus, the primordial black hole with $M_{\rm BH} \lesssim 0.2 M_{\odot}$ may experience the superradiant instability before the recombination of the hydrogen: $z\gtrsim 1100$. 
%For the purpose of constraining PBH abundance, we focus on the redshift $z\lesssim 2\times 10^6$ since for higher redshift the extra energy injection of photons due to the superradiance does not affect the CMB blackbody spectrum. 
If there are PBHs with mass of $10^{-8}M_\odot \lesssim M_{\rm BH} \lesssim 0.2 M_\odot$, photon fields grow exponentially at the redshift $10^3\lesssim z\lesssim 2\times 10^6$, and they can affect the cosmic microwave background (CMB) blackbody spectrum. Thus, PBH abundance with this mass range may be severely constrained.\footnote{
	The lifetime of the PBH through the Hawking radiation is $\tau_{\rm HR} \sim 3\times 10^{72}\,{\rm sec}\,(M_{\rm BH}/M_\odot)^3$. For the PBH mass range of our interest, we can neglect the effect of Hawking radiation. 
}
In the following we mainly focus on the case of $M_{\rm BH}\lesssim 0.2M_{\odot}$.

\begin{figure}[t]
\begin{center}
\includegraphics[scale=0.5]{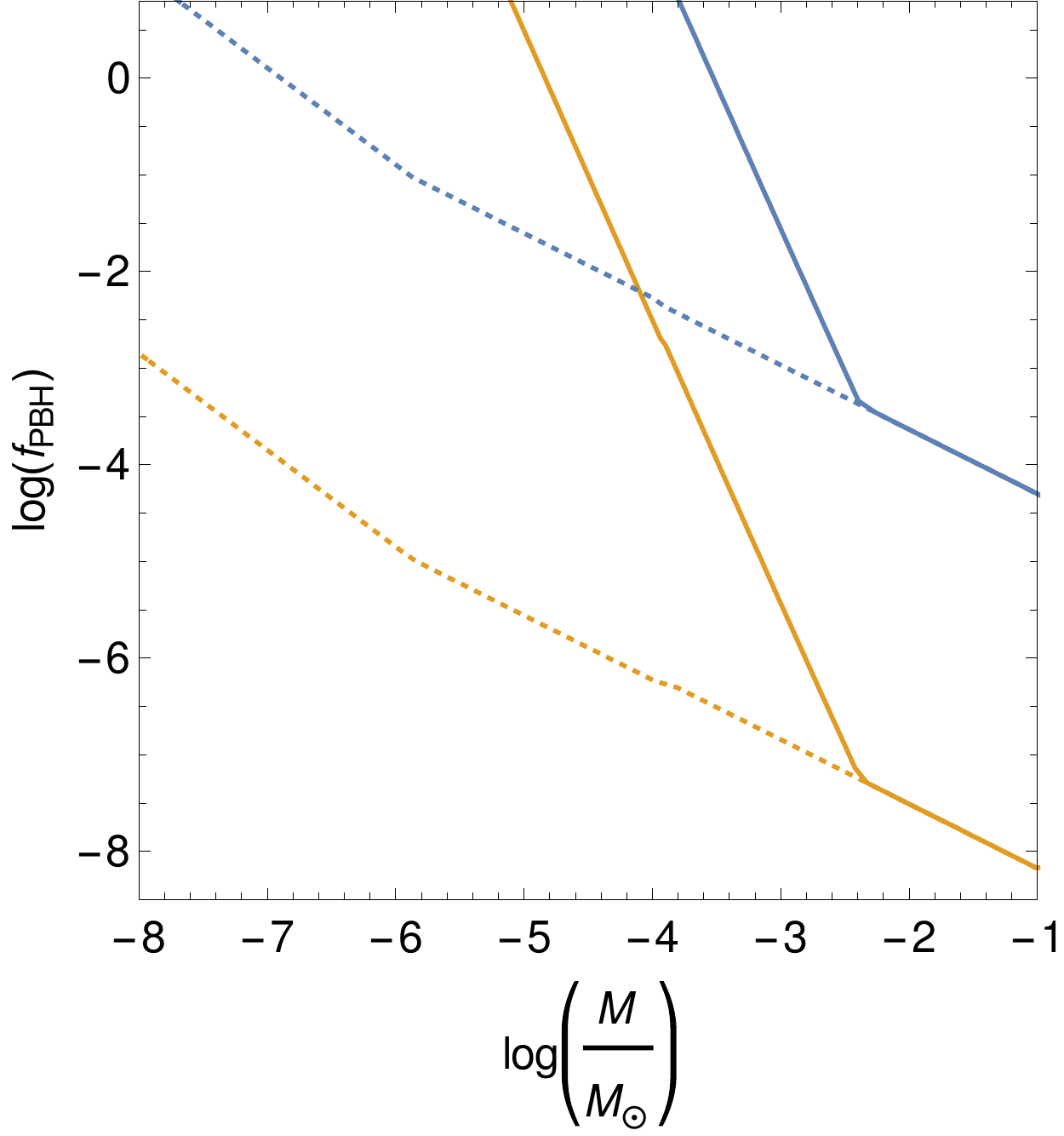}
\includegraphics[scale=0.5]{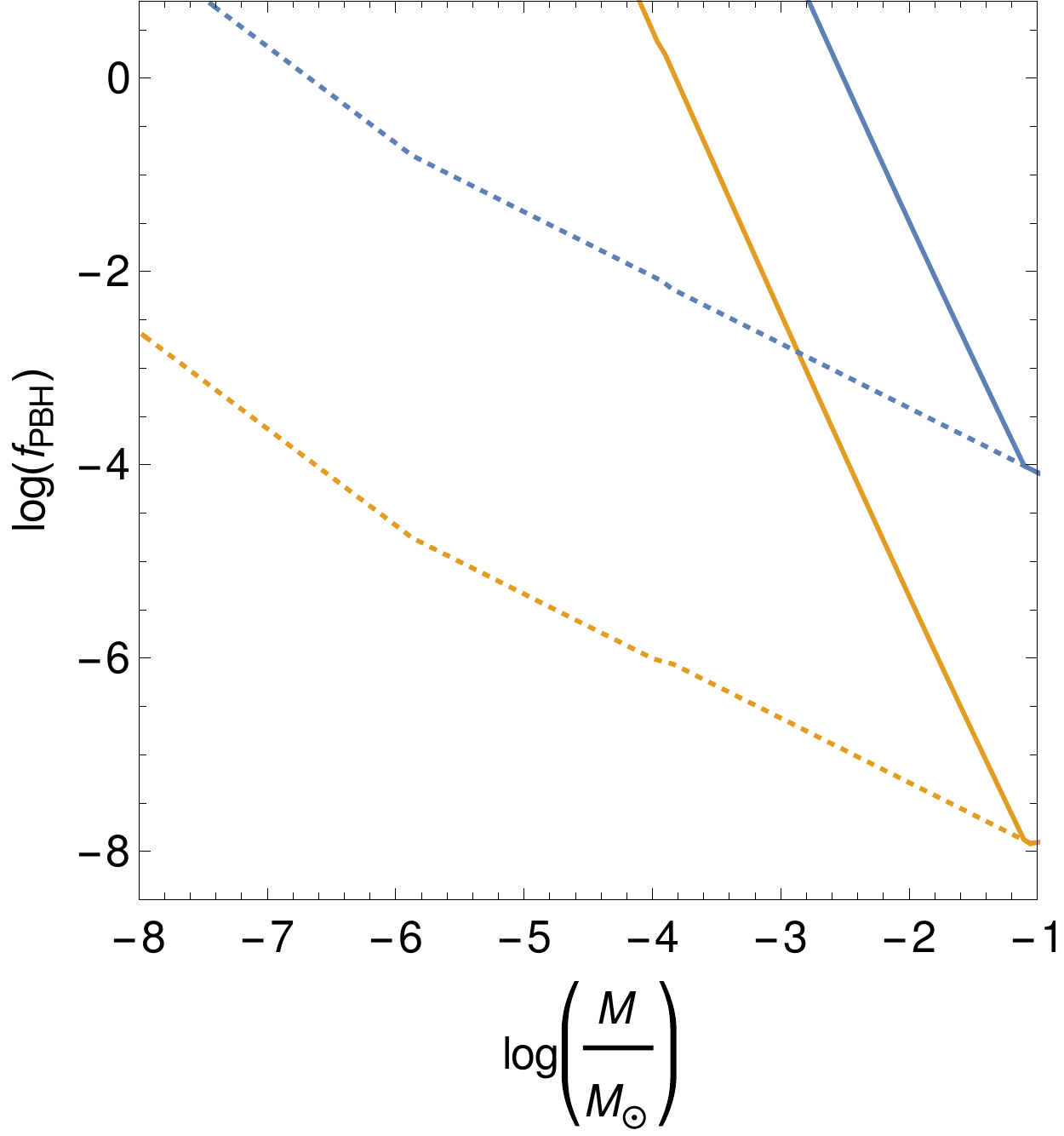}
\end{center}
\caption{
 Constraint on the PBH abundance $f_{\rm PBH}$ as a function of PBH mass. The blue and orange lines are the constraints from COBE and future PIXIE experiments, respectively. The solid and dotted lines are for the constraint with and without considering the photon amplitude saturation by the Schwinger pair production, respectively. For $M \lesssim 10^{-6}M_\odot$, we have used the same assumption as Ref.\,\cite{Pani:2013hpa}. We have taken $(\tilde a, f_{\tilde a})=(1,1)$ in the left panel and $(\tilde a, f_{\tilde a})=(0.6,10^3)$ in the right panel.
}
\label{fig:PBH}
\end{figure}

Now, we include the effect of the nonlinearity, which may suppress the efficiency of the instability. We discuss the Schwinger pair production. The Schwinger effect is reviewed in Appendix~\ref{sec:Sch}.
First let us compare typical time scales. The Hubble time scale is given by
\begin{align}
	\tau_{\rm Hub} = H^{-1} \simeq
	\begin{cases}
	\displaystyle 5\times 10^{9}\,{\rm sec}\,\left(\frac{10^5}{1+z}\right)^2
	 & {\rm for}~~~z\gg z_{\rm eq},\\
	 \displaystyle 8\times 10^{14}\,{\rm sec}\,\left(\frac{10^2}{1+z}\right)^{3/2} & {\rm for}~~~1\ll z\ll z_{\rm eq}
	 \end{cases}
\end{align}
where $z_{\rm eq} \sim 3\times 10^3$ is the redshift at the matter-radiation equality. The superradiant instability time scale is given by
\begin{align}
	\omega_I^{-1} \sim 10^3\,GM_{\rm BH} f_{\tilde a} \sim 5\times 10^{-3}\,{\rm sec}\,f_{\tilde a}\left( \frac{M_{\rm BH}}{M_{\odot}} \right),
\end{align}
where we have introduced a factor $f_{\tilde a}$ that represents the efficiency of the superradiant instability, which is a steep function of the black hole spin $\tilde a$.
For the $(\ell,j)=(0,1)$ mode, it takes $f_{\tilde a} \sim 1$ for $\tilde a\simeq 1$ and $f_{\tilde a} \sim 10^{3}$ for $\tilde a \simeq 0.6$~\cite{Dolan:2018dqv}.
Thus, the instability time scale can be much shorter than the Hubble time scale for the PBH mass of our interest and the photon field grows rapidly.
On the other hand, as explained in the Appendix~\ref{sec:Sch}, the photon energy density around the black hole is saturated at $\rho_A^{\rm max} \equiv \widetilde{\mathcal A}_{\rm max}^2(\pi m_e^2/e)^2$ due to the Schwinger effect, where $ \widetilde{\mathcal A}_{\rm max} \sim 0.05$.
Thus, the time scale of losing $\mathcal O(1)$ fraction of the black hole spin is estimated as
\begin{align}
	\tau_{\rm NL} \sim \frac{J_{\rm BH}}{\dot J_{\rm NL}}\sim \frac{a M_{\rm BH} \omega_p}{\omega_I \rho_A^{\rm max} \mathcal V} \sim 6\times 10^{8}\,{\rm sec}\,\left(\frac{f_{\tilde a}\tilde a \alpha^7}{\widetilde{\mathcal A}_{\rm max}^2}\right)\left( \frac{M_\odot}{M_{\rm BH}} \right).
\end{align}
It can be much longer than the Hubble time scale. Still, however, the gradual energy extraction from the PBH happens. In one Hubble time, the fraction of energy extracted from one PBH is estimated as\footnote{
	Since the plasma frequency changes by $\mathcal O(1)$ after one Hubble time and the instability time scale is a very steep function of $\omega_p$, we can approximate that the instability lasts for about one Hubble time during which $\omega_p GM_{\rm BH} \sim \alpha$.
}
\begin{align}
	f_{\rm ext} \sim \tilde a\alpha\frac{\Delta J_{\rm BH}}{J_{\rm BH}} \sim 
	{\rm min}\left[\tilde a\alpha,~5\times 10^5\, \left(\frac{\widetilde{\mathcal A}_{\rm max}^2}{f_{\tilde a}\alpha^{22/3}}\right)\left(\frac{M_{\rm BH}}{M_\odot}\right)^{7/3}\right].
	%\sim 20\,a^{-1}\, \left(\frac{M_{\rm BH}}{M_\odot}\right)^{7/3},
\end{align}
for $z_M \gg z_{\rm eq}$, where we have substituted $z=z_M$ (\ref{z_M}) assuming $\omega_p GM_{\rm BH} = \alpha$.
Therefore, for $M_{\rm BH}\ll 0.1M_\odot$, we have $f_{\rm ext} \ll \tilde a\alpha$ and the energy extraction due to the superradiant instability is much less efficient than the estimate given in \cite{Pani:2013hpa}.
The extracted energy is liberated in the form of electron-positron pair and they are expected to be mildly relativistic. 
Thus, they affect the CMB spectrum through the so-called $\mu$- or $y$-distortion. For the injection around $10^5\lesssim z \lesssim 2\times 10^6$ the distortion may be characterized by the $\mu$ parameter, which is given by $\mu \simeq 1.4 \delta \rho_r/ \rho_r$ while for the injection around $10^3\lesssim z \lesssim 10^5$ it is characterized by the Compton $y$ parameter, which is given by $y=\delta \rho_r/(4\rho_r)$. The COBE FIRAS experiment puts upper bound on these parameters as $\mu < 9\times 10^{-5}$ and $y< 1.5\times 10^{-5}$~\cite{Fixsen:1996nj}.
In either case, the injection of the radiative energy $\delta \rho_r/\rho_r$ is severely constrained. 
In the present case, we have
\begin{align}
	\frac{\delta\rho_r}{\rho_r} &\simeq f_{\rm ext} f_{\rm PBH}\frac{\rho_{\rm DM}}{\rho_r} \simeq f_{\rm ext} f_{\rm PBH}\left(\frac{1+z_{\rm eq}}{1+z_M}\right)\\
	&\sim 
	\begin{cases}
	\displaystyle 7 f_{\rm PBH}\,\tilde a \alpha^{1/3}\left(\frac{M_{\rm BH}}{M_\odot}\right)^{2/3} & {\rm for}~~f_{\rm ext} \sim \tilde a \alpha \\
	\displaystyle 4\times 10^6 f_{\rm PBH}\left(\frac{\widetilde{\mathcal A}_{\rm max}^2}{f_{\tilde a}\alpha^8}\right)\left(\frac{M_{\rm BH}}{M_\odot}\right)^3 & {\rm for}~~f_{\rm ext} \ll \tilde a \alpha
	\end{cases},
\end{align}
where $f_{\rm PBH}$ denotes the energy fraction of PBH in the total dark matter density.
Therefore, for $M_{\rm BH} \ll 10^{-3} M_\odot$, the energy injection is too small to affect the CMB blackbody spectrum even if $f_{\rm PBH}=1$. Only the mass range $10^{-3}M_\odot \lesssim M_{\rm BH} \lesssim 0.2 M_\odot$ can be constrained from the COBE FIRAS data.
In future, the PIXIE experiment can reach the sensitivity $\mu \sim 10^{-8}$ and $y\sim 10^{-9}$~\cite{Kogut:2011xw} and hence they may be sensitive to the mass range $10^{-5}M_\odot \lesssim M_{\rm BH} \lesssim 0.2 M_\odot$.
Fig.~\ref{fig:PBH} summarizes the constraint on $f_{\rm PBH}$. We have taken $(\tilde a, f_{\tilde a})=(1,1)$ in the left panel and $(\tilde a, f_{\tilde a})=(0.6,10^3)$ in the right panel. 
The PBH abundance with this mass range is also constrained by Subaru HSC~\cite{Niikura:2017zjd}, MACHO~\cite{Allsman:2000kg}, EROS~\cite{Tisserand:2006zx} and OGLE~\cite{2011MNRAS.416.2949W} experiments at the level of $f_{\rm PBH}\lesssim 10^{-3}$--$10^{-1}$. Thus the COBE FIRAS and PIXIE may give more stringent constraint, but one should notice that it crucially depends on the black hole spin $\tilde a$. If the typical size of the PBH spin parameter is $\mathcal O(0.01)$ or below,  $f_{\tilde a}$ becomes extremely large and the CMB observation would not give a meaningful constraint.

%%%%%%%%%%%%%%%%
%%%%%%%%%%%%%%%%

%%%%%%%%%%%%%%%%%%%%%%%%%%%%%%%%%%%%%%
\subsection{Axion with cosine potential}
%%%%%%%%%%%%%%%%%%%%%%%%%%%%%%%%%%%%%%

Let us consider the axion-like particle $\phi$ with a potential
\begin{align}
	V(\phi) = \mu_\phi^2 f^2 \left[1-\cos\left(\frac{\phi}{f}\right) \right],
\end{align}
where $\mu_\phi$ denotes the axion mass and $f$ is the axion decay constant, which we assume to be smaller than the Planck scale: $f\lesssim M_{\rm Pl}$. The axion mass range $10^{-20}\,{\rm eV}\lesssim \mu_\phi\lesssim 10^{-11}\,{\rm eV}$ causes the superradiant instability for the astrophysical black holes with mass of $10^{9} M_{\odot} \gtrsim M_{\rm BH} \gtrsim M_{\odot}$.

The early stage of the superradiant instability is the same as the free massive scalar. 
The typical time scale at this stage is
\begin{align}
	\omega_I^{-1} \sim 10^7\,\mu_\phi^{-1} \simeq 66\,{\rm sec}\left(\frac{10^{-10}\,{\rm eV}}{\mu_\phi}\right).
\end{align}
Initially, the axion cloud exponentially develops, but the non-linearity becomes important when the axion field value becomes close to $f$. The axion potential energy density is bounded as $\rho_\phi < \mu_\phi^2 f^2$ and it implies that the total angular momenta of the axion cloud is bounded as $J_{\rm cloud} / J_{\rm BH} \lesssim 8\pi f^2/(\tilde{a} \alpha^5 M_{\rm Pl}^2)$.
Thus, for $f\ll \alpha^{5/2} M_{\rm Pl}$, the axion cloud extracts only a tiny fraction of the mass and spin of the black hole within a superradiant time scale. However, the axion nonlinear self-interactions cause the emission of the axion particle. Due to this axion emission, the Kerr black hole gradually loses the mass and spin.
The energy loss rate due to the axion emission rate in the massless approximation is estimated as
\begin{align}
	\dot M_{\rm NL} \simeq \frac{1}{16\pi^2}\int d\Omega \left[\int d^3 x' \frac{\mu_\phi^2}{6f^2}\dot \phi \phi^2(t_{\rm ret},\vec x') \right]^2
	\sim C f^{2},
	\label{MNL_axion}
\end{align}
where $t_\text{ret} \equiv t - |x - x'|$, $x$ is the infinite point in $\Omega$ direction and $C$ is a numerical constant that is independent of the axion mass $\mu_\phi$. We take $C\sim 10^{-3}$ from numerical simulation~\cite{Yoshino:2012kn}. Similarly the angular momentum extraction rate is roughly $\dot J_{\rm NL} \sim \dot M_{\rm NL}/ \mu_\phi$\,\footnote{
	It had been pointed out that burst like phenomena called bosenova might happen repeatedly due to the attractive self interaction of the axion~\cite{Arvanitaki:2010sy,Yoshino:2012kn,Yoshino:2013ofa,Arvanitaki:2014wva}. However, an improved numerical simulation is not so supportive as previous studies and the saturation of the axion field is seen~\cite{Yoshino:2019}. Even if the bosenova happens, the estimation (\ref{MNL_axion}) can be used by modifying the constant $C$ $(\sim 10^{-6})$.  
}.
The spin loss time scale of the black hole is then given by
\begin{align}
	\tau_{\rm NL} \sim \frac{J_{\rm BH}}{\dot J_{\rm NL}} \sim 7\times 10^{11}\,{\rm sec}\times\tilde a \alpha \left( \frac{10^{-3}}{C} \right)
	\left( \frac{M_{\rm BH}}{M_\odot} \right)
	\left( \frac{10^{12}\,{\rm GeV}}{f} \right)^2. 
\end{align}
%%

%%%%%%%%%%%%%%%%
\begin{figure}[t]
\begin{center}
\includegraphics[scale=0.7]{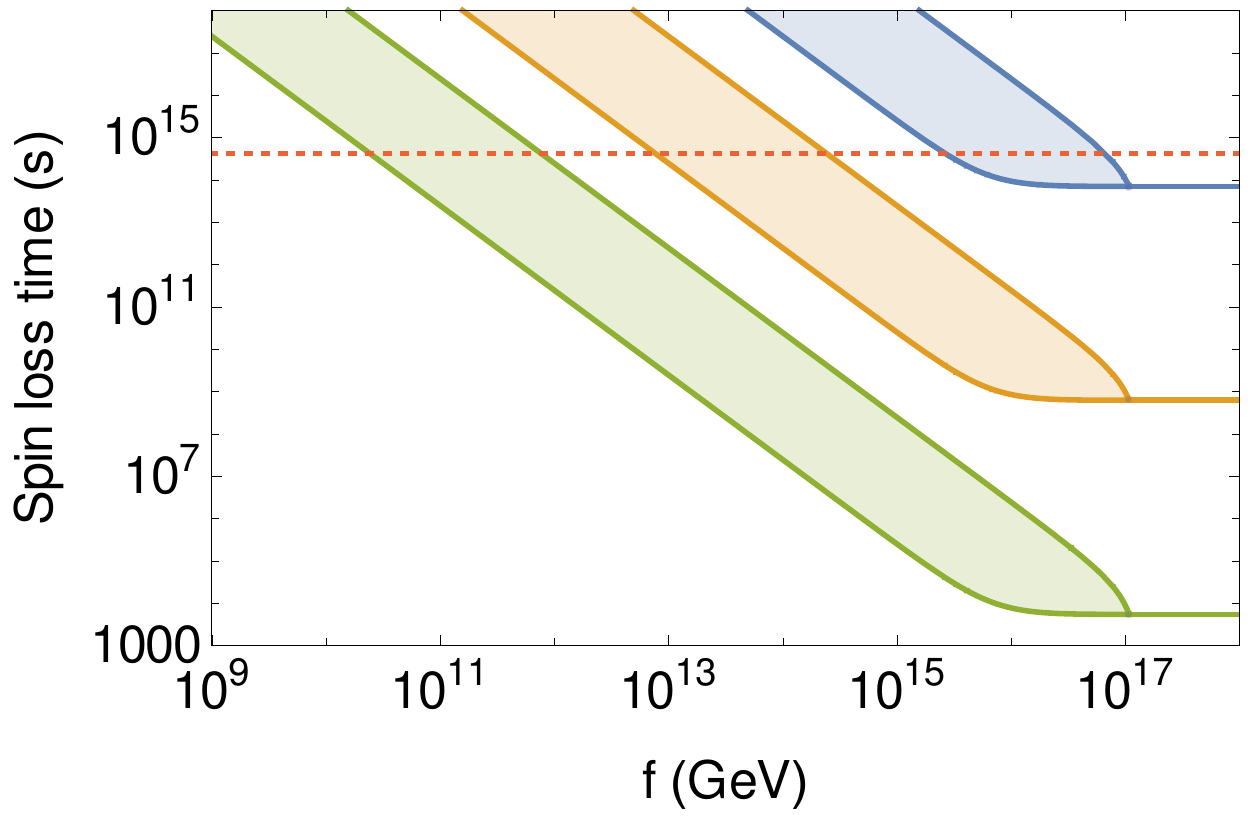}
\end{center}
\caption{
 The relation between the spin loss time by the axion cloud and the axion decay constant. The blue, orange and green bands are for $\mu_\phi = 10^{-20}, 10^{-15}$ and $10^{-10}$\,eV, respectively. The upper (lower) lines of the band correspond to $C=10^{-6}\ (10^{-3})$. The red dotted line shows the accretion time scale, Eq.\eqref{t_acc} with $\epsilon = 0.3$. We take $\tilde a = 1$ and $\alpha = 0.5$. The mass of the Kerr black hole is given as $\alpha / \mu_\phi G$ for each bands, i.e., $M_{\rm BH} / M_{\odot} \sim 10^{10}, 10^5$ and $1$ for the blue, red and green band, respectively.
}
\label{fig:Axion_spin_loss}
\end{figure}
%%%%%%%%%%%%%%%%

We have shown the time for the Kerr black hole to lose $\mathcal O(1)$ of the angular momentum in term of $f$ in Fig.\,\ref{fig:Axion_spin_loss}. As we have discussed, the exponential growth of the axion cloud efficiently extracts the angular momentum for $f \gtrsim 10^{17}\,\text{GeV}$ and the spin loss time is determined by just the superradiance time scale $\tau_{\rm SR}$ (see footnote \ref{footnote:SR}). On the other hand, for much smaller $f$, the energy density of the cloud is saturated and the spin is decreased only linearly with time due to the axion emission. In this regime the spin loss time scale is mainly  determined by $\tau_{\rm NL}$.
The efficiency may depend on $C$, but for some parameter regions, the spin loss rate by the particle emission is faster than the accretion time scale, Eq.\eqref{t_acc}. Thus the observation of high spin black holes with these region may put a constraint on the corresponding axion mass even if the exponential growth ceases due to the nonlinearity. In particular, for $10^{16}\,\text{GeV} \lesssim f \lesssim 10^{17}\,\text{GeV}$, the particle creation process time scale is as fast as the superradiance time scale for $C \sim 10^3$.
On the other hand, if the accretion time scale is shorter than the spin down time scale, the superradiance does not much affect the black hole evolution. 
%Typically the accretion time scale is much longer than $\tau_{\rm NL}$ and hence the accretion does not much affect the discussion above.

%%%%%%%%%%%%%%%%%%%%%%%%%%%%%%%%%%%%%%
\subsection{Hidden photon with Higgs mechanism}
%%%%%%%%%%%%%%%%%%%%%%%%%%%%%%%%%%%%%%

Next let us consider the black hole superradiance with the light hidden photon field. We assume that the hidden photon mass is generated by the Higgs mechanism.\footnote{
	The effect of Higgs interaction on the vector boson superradiance was briefly mentioned in Ref.~\cite{Baryakhtar:2017ngi}.  
} The relevant Lagrangian is
\begin{align}
	\mathcal L =  -\frac{1}{4}F_{\mu\nu}F^{\mu\nu} + |D_\mu\Phi|^2-V(\Phi) ,
\end{align}
where $\Phi$ denotes the Higgs field, $D_\mu \Phi = \partial_\mu\Phi-ig A_\mu \Phi$ and $g$ is the gauge coupling constant. The Higgs potential $V(\Phi)$ is arranged so that the Higgs field obtains a VEV $|\Phi|=v/\sqrt{2}$. By using the gauge U(1) degree of freedom, one can take the unitary gauge such that the Higgs field is expanded as $\Phi = (v + \sigma)/\sqrt{2}$ where $\sigma$ is the radial fluctuation of the Higgs and the Goldstone mode is gauged away. The Lagrangian is then given by
\begin{align}
	\mathcal L =  -\frac{1}{4}F_{\mu\nu}F^{\mu\nu} +\frac{1}{2}(\partial_\mu \sigma)^2-V(\sigma) +\frac{1}{2}g^2(v+ \sigma)^2 A_\mu A^\mu.
\end{align}
In the vacuum, $\sigma=0$, the hidden photon has a mass of $\mu_A = gv$. 
In the limit of heavy $\sigma$, we can neglect the dynamics of $\sigma$ and we are left with just a massive hidden photon theory, as also realized in the Stuckelberg mechanism. However,  there are still nontrivial phenomenological effects by $\sigma$ on the superradiant instability of the vector boson unless $\sigma$ is infinitely heavy, as explained below. Since we are interested in the very light hidden photon with mass of $\mu_A \lesssim 10^{-11}\,{\rm eV}$, we focus on the case of $\mu_\sigma \gg \mu_A$ where $\mu_\sigma$ is the $\sigma$ mass.
On the other hand, assuming the perturbativity of the Higgs self coupling, we have an inequality $\mu_\sigma \lesssim v = \mu_A/g$. Thus we need very small $g$ for satisfying $\mu_\sigma \gg \mu_A$.

Let us suppose that $\mu_A=gv$ satisfies the superradiant condition. Then, the vector field is amplified around the rotating black hole and $\sigma$ gets an additional potential term of $g^2(v+\sigma)^2\left<A_\mu A^\mu\right>/2$ due to the finite density effect.
If the physical Higgs mass is large enough, i.e., $\mu_\sigma \gg \mu_A$, one can neglect the dynamics of $\sigma$ and integrate out it. For concreteness, we take the Higgs potential as $V(\Phi)=\lambda (|\Phi|^2-v^2/2)^2$. In this case, we have $\mu_\sigma^2=2\lambda v^2$.
Taking account of the vector background, one can find the extrema of the effective potential at
\begin{align}
	\sigma = -v~~~{\rm and}~~~v\left(-1\pm \sqrt{1-\frac{X}{X_{\rm max}}} \right).  \label{sigma_min}
\end{align}
where we have defined $X\equiv -A_\mu A^\mu$ and
\begin{align}
    X_{\rm max} \equiv \frac{\lambda v^2}{g^2} = \frac{\mu_\sigma^2}{2g^2}.
\end{align}
For $X < X_{\rm max}$, there are minima represented by the second solutions of (\ref{sigma_min}) and $\sigma$ tracks this temporal minimum of the potential. For $X>X_{\rm max}$ these solutions disappear and $\sigma=-v$ becomes a minimum, which means the symmetry restoration.
In the following, we assume $X < X_{\rm max}$.
The resulting effective ``potential'' of the vector field is obtained by substituting the second solution of (\ref{sigma_min}) into the original potential and given by
\begin{align}
	V_{\rm eff}(A) = \frac{1}{2}\mu_A^2 X \left(1-\frac{X}{2X_{\rm max}} \right).  \label{VeffA}
\end{align}
This nonlinear self-interactions of the vector boson has appeared after integrating out $\sigma$.
%It becomes effectively flat at large $X$, since $|\Phi| = (v+\sigma)/\sqrt{2}$ becomes smaller and the vector boson becomes lighter. For $g^2 X \gtrsim \mu_\sigma^2$ the finite density effect is so strong that the symmetry is restored: $\Phi =0$ or $v+\sigma=0$.
This implies that there is an upper bound on the vector field $X$ above which the backreaction to the Higgs field becomes important and the symmetry is restored. In a realistic setup, the superradiant instability is expected to effectively stops when the nonlinearity of the vector boson becomes important before the symmetry restoration happens, similarly to the case of axion. In any case, the upper bound is roughly estimated as $X \sim X_{\rm max}$.
Thus, the energy density of the vector boson cloud around a rotating black hole is bounded as $\rho_A^{\rm max} \sim \mu_A^2 \mu_\sigma^2/g^2$. Then, the ratio of the total angular momenta of the cloud and the black hole is
\begin{align}
	\frac{J_{\rm cloud}}{J_{\rm BH}} \lesssim \frac{\mathcal V \rho_A^{\rm max}}{\tilde{a}\alpha M_{\rm BH}} \sim \frac{\pi \left<X\right>_{\rm max}}{8\pi\tilde a\alpha^5 M_{\rm Pl}^2}
	\sim \frac{\mu_\sigma^2}{8\tilde a\alpha^5 g^2 M_{\rm Pl}^2},
\end{align}
where $\mathcal V$ denotes the effective volume of the cloud and we took $\mathcal V \simeq\pi (\alpha \mu_A)^{-3}$ (see Appendix~\ref{sec:Sch}). Therefore, for $\mu_\sigma\ll \alpha^{5/2} g M_{\rm Pl}$, the total angular momenta of the vector cloud is much smaller than that of the central black hole and hence the superradiance cannot take a substantial fraction of the black hole energy and spin away.

Here is one remark. In the above discussion $X\simeq -(A_0)^2 + (A_i)^2$ is assumed to grow through the superradiance. It is justified as follows. In a pure massive vector field theory it is known that the vector field automatically satisfies the Lorentz condition $D_\mu A^\mu=0$ through the equation of motion. One can show that the same is true also for a theory with effective vector potential (\ref{VeffA}). Since the typical time variation scale of the vector cloud is $\mu_A^{-1}$ while the spatially varying scale is $(\mu_A\alpha)^{-1}$, we should have $(A_i)^2 > (A_0)^2$ to satisfy the Lorentz condition and hence we can take $X \sim (A_i)^2$.

As in the case of axion, still there may be energy extraction processes from the system of black hole and vector boson cloud. 
Note that the oscillating $A_\mu$ field cannot induce particle production of $\sigma$ in our setup since $g^2 X\mu_A^2 \gtrsim \mu_\sigma^4$, the condition we discuss in the next subsection, is not satisfied.
However, there are effective self-interactions of the vector field as expressed in (\ref{VeffA}). It induces the emission of the vector boson and extracts the energy and angular momentum of the system. Similarly to the axion case, we can estimate the emission rate as
\begin{align}
	\dot M_{\rm NL} \simeq \frac{1}{16\pi^2} \int d\Omega \left[\frac{\partial}{\partial t}\int d^3 x' \frac{2g^2\mu_A^2}{\mu_\sigma^2} X A_i (t_{\rm ret},\vec x') \right]^2 \simeq C \frac{\mu_\sigma^2}{g^2},
\end{align}
where $C$ is a numerical constant independent of the vector boson mass $m_A$. A detailed numerical simulation is required to find a value of $C$. The energy/spin loss time scale of the black hole is then given by
\begin{align}
	\tau_{\rm NL} \sim \frac{J_{\rm BH}}{\dot J_{\rm NL}} \sim 7\times 10^{11}\,{\rm sec}\times \tilde a\alpha\left( \frac{10^{-3}}{C} \right)
	\left( \frac{M_{\rm BH}}{M_\odot} \right)
	\left( \frac{10^{12}\,{\rm GeV}}{\mu_\sigma/g} \right)^2,
\end{align}
which significantly depends on the value of $\mu_\sigma/g$.

%%%%%%%%%%%%%%%%
\begin{figure}[t]
\begin{center}
\includegraphics[scale=0.7]{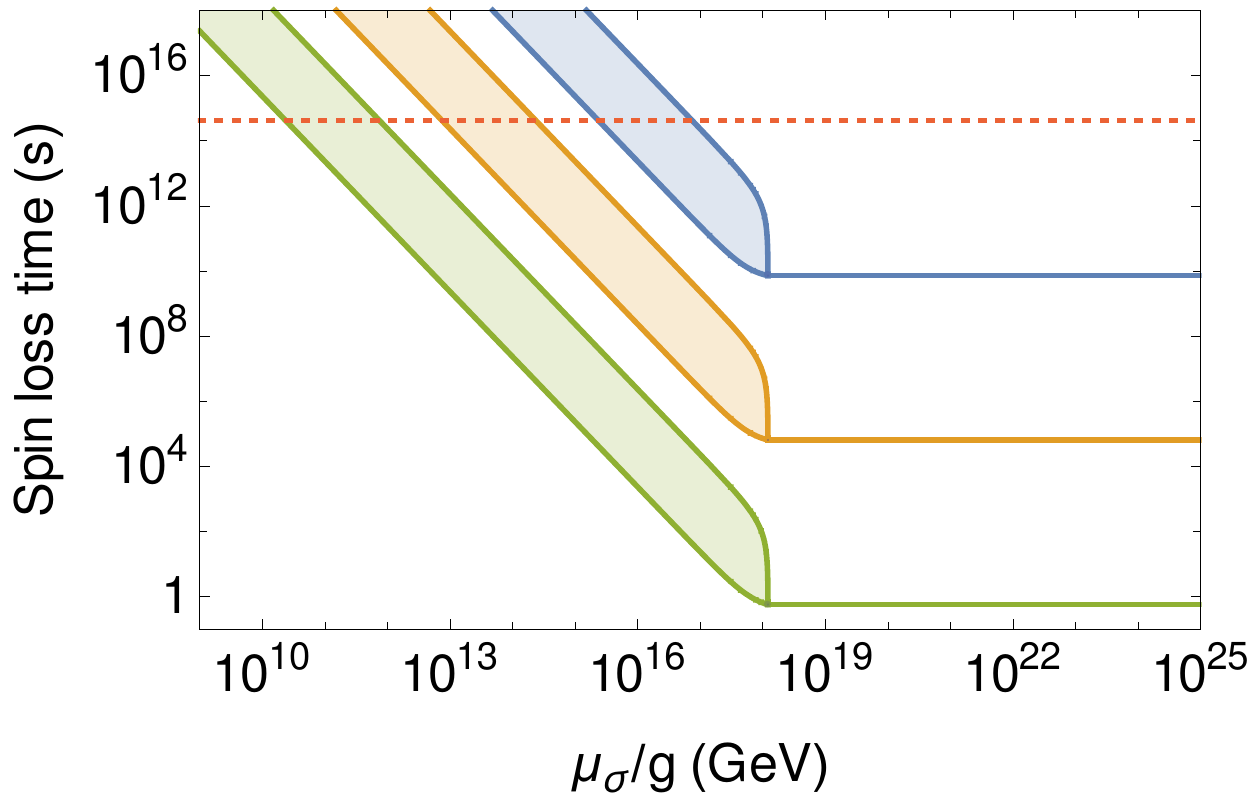}
\end{center}
\caption{
 The relation between the spin loss time by the hidden photon cloud and $\mu_\sigma / g$. The blue, orange and green bands are for $\mu_A = 10^{-20}, 10^{-15}$ and $10^{-10}$\,eV, respectively. The upper (lower) lines of the band correspond to $C=10^{-6}\ (10^{-3})$. The red dotted line shows the accretion time scale, Eq.\eqref{t_acc} with $\epsilon = 0.3$. We take $\tilde a = 1$ and $\alpha = 0.5$. The mass of the Kerr black hole is given as $\alpha / \mu_A G$ for each bands, i.e., $M_{\rm BH} / M_{\odot} \sim 10^{10}, 10^5$ and $1$ for the blue, red and green band, respectively. 
}
\label{fig:HP_spin_loss}
\end{figure}
%%%%%%%%%%%%%%%%

We have again shown the relation between the spin loss time and the characteristic scale, $\mu_\sigma / g$, in Fig.\,\ref{fig:HP_spin_loss}. As we have discussed, the exponential growth eventually stops for $\mu_\sigma / g \gtrsim 10^{18}\,\text{GeV}$. However, as is the case for the axion, in some parameter regions, the nonlinear particle emission process is faster than the accretion and the substantial spin of the black hole is extracted. Thus, such parameter regions can be constrained from observations.

%On the other hand, for $m_\sigma \gg g M_{\rm Pl}$, the theory is not distinguished from a purely massive vector theory without the Higgs and hence the vector cloud can grow until it extract a substantial fraction of central black hole spin and hence the existence of such a light vector boson is constrained from the observation of spinning black hole.

%%%%%%%%%%%%%%%%%%%%%%%%%%%%%%%%%%%%%%
\subsection{Scalar with four-point interaction}
\label{sec:pp_scalar}
%%%%%%%%%%%%%%%%%%%%%%%%%%%%%%%%%%%%%%

Finally, let us discuss a scenario where the superradiant degree of freedom is a scalar boson, $\phi$, which interacts with another scalar particle, $\chi$. Ignoring the other interactions, the Lagrangian is 
\begin{align}
\mathcal{L} = \frac{1}{2}(\partial\phi)^2 + \frac{1}{2}(\partial\chi)^2 - \frac{\mu_\phi^2}{2}\phi^2 - \frac{\mu_\chi^2}{2}\chi^2 - \frac{g^2}{2}\phi^2\chi^2.
\end{align}
In the following, we only consider the case of $\mu_\chi\gg \mu_\phi$ for simplicity.
Note that the $\phi^2\chi^2$ interaction necessarily introduces the effective self-interaction of $\phi$ as
\begin{align}
	V_{\rm eff}(\phi) = \frac{(g^2\phi^2+\mu_\chi^2)^2}{64\pi^2} \left[\log \left(\frac{g^2\phi^2+\mu_\chi^2}{\Lambda^2}\right) - \frac{3}{2}\right],
	\label{V_CW}
\end{align}
where $\Lambda$ denotes the renormalization point.\footnote{
	It should be understood that the mass of $\phi$ around the origin $\phi=0$ is renormalized to be $\mu_\phi$ after summing up the bare mass and that arises from (\ref{V_CW}). Similarly, the four point $\phi$ self coupling is renormalized to be zero around the origin.
}
Thus the $\phi$ potential becomes effectively quartic for $|\phi| \gtrsim \phi_{\rm NL}\equiv 8\pi \mu_\phi / g^2$.

%Note that the self-interaction necessarily emerges from the interaction, but we will discuss the effect later and omit it here.
First, let us suppose that the $\phi$ potential is well approximated by the quadratic one: $V=\mu_\phi^2\phi^2/2$.
%We could again integrate $\chi$ out to obtain the effective action and discuss the vacuum-to-vacuum amplitude\,\cite{Schwinger-Preheating}, but alternatively, we solve the classical field equation of motion directly. 
Assuming that $\phi = \phi_0 \sin \mu_\phi t$ inside the superradiance cloud, the equation of motion for $\chi_k(t)$, the $k$-mode of $\chi$ field, is
\begin{align}
\label{eq:Chi_EOM}
    \ddot\chi_k + (\mu_\chi^2 + k^2 + g^2 \phi_0^2 \sin^2 \mu_\phi t)\chi_k = 0.
\end{align}
Although the background geometry is not flat, the Fourier decomposition is justified as far as we are interested in the short wavelength modes.
This is the Mathieu equation that is analyzed in detail in the context of reheating after inflation\,\cite{Dolgov:1989us,Traschen:1990sw,Shtanov:1994ce,Kofman:1994rk,Kofman:1997yn}. Let us shortly review what happens in the limit where $\phi$ is spatially homogeneous. 
First note that in the small $\phi$ amplitude regime essentially no particle production happens since $\mu_\chi\gg \mu_\phi$ and the perturbative particle production is kinematically forbidden.
For the large amplitude regime, $g \phi_0 \mu_\phi \gg \mu_\chi^2$, there are $\chi$ modes satisfying
\begin{align}
	g \phi_0 \mu_\phi &\gtrsim \mu_\chi^2 + k^2
\end{align}
and the $\chi$ particle with such momenta $k$ exponentially grows through the so-called broad parametric resonance~\cite{Kofman:1994rk, Kofman:1997yn}.
In this regime $\mathcal{O}(1)$ particles per volume $\Delta \mathcal V \sim k_*^{-3}$ are produced within time duration of $\Delta t \sim \mu_\phi^{-1}$, where $k_*^2\equiv g\phi_0 \mu_\phi- \mu_\chi^2$. 
Ignoring the self-interaction of $\phi$ at this stage is justified if
\begin{align}
	g \mu_\chi^2 \lesssim 8\pi \mu_\phi^2.   \label{phi_nonlinear}
\end{align}
For a while, we assume that this inequality is satisfied. Otherwise, the $\phi$ potential would be dominated by the effective quartic one $V\sim (g^4\phi^4/64\pi^2) \log(\phi^2)$ before the particle production is switched on.

Now let us adopt a similar analysis to the boson cloud around a black hole. At the first stage the interaction term is negligible and $\phi$ cloud begins to grow due to the superradiant instability. When the amplitude reaches around $g\phi_0 \mu_\phi \sim \mu_\chi^2$, the $\chi$ particle production begins to be efficient. 
% Note that, in the beginning the wavelength of the produced $\chi$ particle is expected be so long that we cannot rely on the homogeneous field approximation and the analysis will become much more complicated. However, after a few superradiant instability time scale, the typical wavelength of the produced $\chi$ particle becomes much smaller than the boson cloud radius and the homogenous field approximation becomes valid. Below we make use of this approximation.

It should be noticed that the produced $\chi$ particles are relativistic at the very instant of their production, which happens at the time interval $\Delta t \sim \sqrt{k^2+\mu_\chi^2}/(g\phi_0 \mu_\phi)$ around when $\phi$ passes through $\phi=0$, but they soon become non-relativistic as $\phi$ increases again. It implies that the most of the created particles cannot escape from the gravity of black hole, since a particle must be at least semi-relativistic during the time interval longer than $GM_{\rm BH} \sim \mu_\phi^{-1}$ in order to escape.
Therefore, through the particle production process, the $\chi$ cloud appears in association with superradiant $\phi$ cloud. Their time evolution is described by
\begin{align}
	&\dot M_{\rm cloud}^{(\phi)} = 2\omega_I M_{\rm cloud}^{(\phi)} - \dot M_{\rm prod},\\
	&\dot M_{\rm cloud}^{(\chi)}= \dot M_{\rm prod}, 
\end{align}
where $\dot M_{\rm prod}$ denotes the energy transfer rate due to particle production, which is basically an increasing function of $\phi_0$.\footnote{
	After time average over one $\phi$ oscillation $\mu_\phi^{-1}$, we may have $\dot M_{\rm prod}\sim g\mu_\phi \phi_0 k_*^3 \mathcal V$.
}
Thus, it is expected that the growth of $\phi$ cloud stops when $\dot M_{\rm prod}$ becomes comparable to the superradiant growth rate $2\omega_I M_{\rm cloud}^{(\phi)}$. $\chi$ cloud still continues to grow and eventually the backreaction of $\chi$ to the $\phi$ potential becomes important, when the $\chi$ and $\phi$ energy density become comparable. Then, the $\chi$ particle production is terminated.

If the inequality (\ref{phi_nonlinear}) is inverted, the effective $\phi^4$ potential becomes important before the particle production process becomes efficient. In this case, the superradiance is expected to stop at $\phi_0 \lesssim \phi_{\rm NL}$.
Therefore, in either case, the interaction term tends to make the superradiant instability inefficent.
A precise estimation is difficult because of the nontrivial configuration of the both clouds and nonlinearity, but it is a reasonable expectation that the growth of the cloud stops when the particle production becomes efficient or the nonlinearity of the potential becomes effective.

Finally, we briefly comment on the case of interaction with Fermion,
\begin{align}
\mathcal{L} = \frac{1}{2}(\partial\phi)^2 -\frac{\mu_\phi^2}{2}\phi^2 + i\bar{\psi}\slashed{\partial}\psi - \mu_\psi \bar{\psi}\psi - y\phi\bar{\psi}\psi.
\end{align}
The broad parametric resonance again creates $\psi$ particle out of the $\phi$ background. The difference is that the parametric resonance does not grow up exponentially unlike the scalar interaction because of the Pauli's exclusion principle\,\cite{Greene:1998nh,Greene:2000ew,Peloso:2000hy}.
%In our system, the created particle may escape the black hole system and evade the Pauli's exclusion principle although as we have discussed the velocity of the produced particle is small and the escape rate is expected not to be large. In order to estimate the effective energy leakage, we need the numerical analysis.
On the other hand, the $\psi$ loop again generates the non-linear self interaction for $\phi$. This may make the superradiance inefficient.

%%%%%%%%%%%%%%%%%%%%%%%%%%%%%%%%%%%%%%
\section{Conclusions and Discussion} \label{sec:dis}
%%%%%%%%%%%%%%%%%%%%%%%%%%%%%%%%%%%%%%

Recently, the black hole superradiance phenomena have been drawing lots of attention as a probe of ultralight boson fields. The most previous studies focused on the case where the boson is a free field and there have been much progress on the understanding of the physics of superradiance and its phenomenological implications.

In this paper, we have discussed several effects of nonlinear interactions of the boson field on the superradiance and the resultant evolution of black holes. Although it is difficult to precisely calculate the evolution of the boson cloud and the black hole due to the nonlinearity, we can still make reasonable estimates for the nonlinear effects.
One of the key effects is the saturation of the field amplitude at which the nonlinearity becomes important. It has two possible origins: the modification of the scalar potential itself and the effect of particle production.
We partly relied on the fact that the numerical simulation of the self-interacting axion cloud shows a saturation of the field value around which the nonlinear effect becomes important~\cite{Yoshino:2012kn,Yoshino:2019}. It is not clear what happens for the general form of the nonlinear scalar potential, but it is unlikely that the cloud continues to grow even if the particle production becomes very efficient. We need further studies on this point.
The other key effect is that the nonlinear interactions lead to the emission of high momentum particles from the boson cloud, which extracts energy and angular momentum of the cloud. Even if the field amplitude is saturated as explained above, there is a gradual energy loss process.

Taking these effects into account, we have considered the evolution of the black hole and surrounding boson cloud for some concrete examples. The standard model photon can experience the superradiant instability since it has a plasma mass and it can satisfy the superradiant condition if there are PBHs in the early universe. However, it necessarily causes the Schwinger pair production as the photon field grows and there is an upper bound on the efficiency of the energy injection from PBHs to the plasma. We have shown that the constraint on the PBH abundance may be much weaker than the previous estimate.

The axion with cosine potential is also considered. This case is already studied numerically~\cite{Yoshino:2012kn,Yoshino:2019}. Our whole picture, i.e., the saturation of the field value and the extraction of energy and angular momentum, is roughly consistent with the numerical study.

We have also discussed the light hidden photon whose mass comes from the Higgs mechanism. It is shown that the growth of the hidden photon due to the superradiant instability modifies the Higgs potential so that the configuration of the Higgs expectation value around the black hole becomes nontrivial. In the limit of heavy Higgs, we effectively obtain a theory of the self-interacting hidden photon, which is a bit similar to the axion.
Depending on the value of $\mu_\sigma/g$, the production of the hidden photon can be so inefficient that there are essentially no observable consequences. Conversely, we can constrain $\mu_\sigma/g$ from observations.

Our study may not go beyond rough order-of-magnitude estimations, but it shows several drastic effects on the boson cloud and black holes and their observational consequences.
Since there are a priori no reasons to expect that these ultralight bosons are free massive fields, it is essential to understand nonlinear effects precisely for the purpose to prove a nature of ultralight bosons, although they require detailed  numerical simulations. We leave these issues to future works.

%%%%%%%%%%%%%%%%%%%%%%%%%%%%%%%%%%%%%%%%%%%%
\acknowledgments
%%%%%%%%%%%%%%%%%%%%%%%%%%%%%%%%%%%%%%%%%%%%

We thank Hirotaka Yoshino for useful discussion.
This work was supported by the Grant-in-Aid for Scientific Research C (No.18K03609 [KN]) and Innovative Areas (No. 16H06490 [HF], No.15H05888 [KN], No.17H06359 [KN]).
HF was supported by the Director, Office of Science, Office of
High Energy Physics of the U.S. Department of Energy under the
Contract No. DE-AC02-05CH11231.

%%%%%%%%%%%%%%%
\appendix
%%%%%%%%%%%%%%%

%%%%%%%%%%%%%%%%%%%%%%%%%%%%%%
\section{Schwinger pair production}  \label{sec:Sch}
%%%%%%%%%%%%%%%%%%%%%%%%%%%%%%

%%%%%%%%%%%%%%%%%%%%%%%%%%%%%%
\subsection{Schwinger pair production rate} 
%%%%%%%%%%%%%%%%%%%%%%%%%%%%%%

Here, we discuss a superradiant instability induced by a massive vector boson, $A_\mu$, with a current interaction between a Dirac Fermion with charge $-1$, $\psi$, which we call as the electron.
The Lagrangian for $A_\mu$ is
\begin{align}
    \mathcal{L} = -\frac{1}{4}F_{\mu\nu} F^{\mu\nu} + \frac{\mu_A^2}{2} A^2 - e A_\mu J^\mu + i\bar{\psi}\slashed{\partial}\psi - m_e \bar{\psi}\psi,
\end{align}
where $F_{\mu\nu} \equiv \partial_\mu A_\nu - \partial_\nu A_\mu$, $\mu_A$ is the mass of $A$, $e$ is the matter charge, $J_\mu$ is the electron current and $m_e$ is the electron mass.

The effective Lagrangian is obtained by integrating the electron fields out\,\cite{Schwartz:2013pla,Itzykson:1980rh}.
\begin{align}
    \mathcal{L}_\text{eff} = -\frac{1}{4}F^2 + \frac{\mu_A^2}{2} A^2 - i \text{Tr}\left[\left\langle x |\ln(i\slashed{D} - m_e) |x \right\rangle\right],
\end{align}
where $D_\mu \equiv i\hat{p}_\mu - ieA(\hat{x})_\mu$, $|x\rangle$ is the eigenvector of an infinite dimensional matrix $\hat{x}_\mu$ with an eigenvalue $x_\mu$ corresponding to the spacetime coordinate and $\hat{p}$ is the matrix satisfying $[\hat x_\mu, \hat p_\nu] = -i\eta_{\mu\nu}$.
Ignoring the mass term, for the constant electromagnetic field $F_{\mu\nu}$, we can calculate the matrix element exactly and obtains the so-called Euler-Heisenberg Lagrangian\,\cite{Heisenberg1936}. 
\begin{align}
\label{eq:Euler_Heisenberg}
    \mathcal{L}_{\rm EH} = -\frac{1}{4}F^2 + \frac{\alpha_e}{8\pi} \int_0^\infty ds e^{-sm_e^2 + is\varepsilon} \left[\frac{\text{Re}\cosh(e s X)}{\text{Im}\cosh(e s X)} F_{\mu\nu}\widetilde{F}^{\mu\nu} + \frac{4}{e^2s^2} + \frac{2}{3}F^2\right],
\end{align}
where $\alpha_e\equiv e^2 / 4\pi$, $X\equiv\sqrt{\frac12 F^2 - \frac i2F\widetilde F}$ and $\widetilde{F}_{\mu\nu} \equiv \frac12 \epsilon_{\mu\nu\rho\sigma} F^{\rho\sigma}$.

The effective action, $\Gamma_\text{eff} \equiv \int d^4x \mathcal{L}_{\rm EH}$, governs the vacuum-to-vacuum amplitude for the given constant $F_{\mu\nu}$ background. If the amplitude $\left|\exp(i\Gamma_{\text{eff}})\right|^2$ is smaller than unity, then the other states, the electron and positron pair, emerge from the background. Thus, the discrepancy is the electron pair creation rate.
Because
\begin{align}
    \left|e^{i\Gamma_{\text{eff}}}\right|^2 = e^{-2 \text{Im}\Gamma},
\end{align}
the electron pair production rate per unit volume, $\Gamma_S$, is given as
\begin{eqnarray}
\label{eq:Schwinger_rate}
\Gamma_S(E) = 2 \text{Im}\mathcal{L}_\text{eff} = \frac{\alpha_e E^2}{\pi^2}\sum_{n = 1}^\infty \frac{1}{n^2}\exp\left(-\frac{n\pi m_e^2}{e E}\right).
\end{eqnarray}
Here, the magnetic field is assumed to be zero and the electric field is denoted by $E$. This pair-creation process is called the Schwinger effect\,\cite{Schwinger:1951nm}.

%%%%%%%%%%%%%%%%%%%%%%%%%%%%%%
\subsection{Comparison with superradiance rate} 
%%%%%%%%%%%%%%%%%%%%%%%%%%%%%%

We compare the energy loss rate by the Schwinger effect with the energy extraction by superradiance from a Kerr black hole. 
%We will later justify the use of the Schwinger effect, where the electromagnetic field is constant.
Around the Kerr black hole, the superradiant cloud has a size around $(\alpha \mu_A)^{-1}$. Thus the typical magnitude of the electric field is $E \simeq \mu_A A$, where $A$ is the typical amplitude of the vector boson. Due to the superradiant instability, $A(t)$ grows exponentially with a frequency $\omega_I$: $A(t) \propto \exp(\omega_I t)$. The energy density carried by the vector field is $\rho_A \sim \mu_A^2 A^2/2$. As the vector cloud grows, electron-positron pair is produced through the Schwinger effect, which reduces the energy of the vector cloud. The change of the energy density of the vector boson cloud around the Kerr black hole may be described by
\begin{align}
	&\dot M_\text{cloud} \simeq 2\omega_I M_{\rm cloud} - 2m_e \int \Gamma_S(E)\,d^3 x.
	\label{Mdot_schwinger}
\end{align}
%where $n_{e^+}$ and $n_{e^-}$ are positron and electron number density produced through the Schwinger process and $\Gamma_{\rm ann}$ collectively denotes the reduction rate of electron/positron number density due to the pair annihilation, interaction with plasma, absorption by the black hole, etc. First, for simplicity, we neglect the effect of $\Gamma_{\rm ann}$.
For $E > \pi m_e^2 / e$, the Schwinger pair production process is unsuppressed. In this case one can easily estimate that the Schwinger production rate exceeds the superradaince rate if $\omega_I < \alpha_e m_e/\pi^2$, which is satisfied for the standard model photon and electron.
Therefore, it is expected that the superradiant growth stops at some instant where the Schwinger production rate becomes comparable to the superradiance rate.

Note that the produced electrons and positrons are accelerated by the electric field but they do not give net energy loss of the vector cloud. This is because the electric field $\vec E$ is oscillating with time scale $\mu_A^{-1}$ and correspondingly the velocity of the electron/positron $\vec v$ is also oscillating, but the work done by the electric field is proportional to $\vec E\cdot \vec v$ that becomes zero after time average.
Actually, however, there are number of effects that can reduce the electron/positron energy: the electron-positron pair annihilation, synchrotron emission associated with the magnetic field, interaction with plasma, absorption by the black hole, etc. 
Nevertheless, Eq.~(\ref{Mdot_schwinger}) gives a conservative estimate for the upper bound on the magnitude of the vector boson amplitude. 
As will become clear below, the actual upper bound is not so sensitive to the detailed process because the Schwinger production rate is exponentially dependent on the vector boson amplitude.

\begin{figure}[ht]
\begin{center}
\includegraphics[scale=1.0]{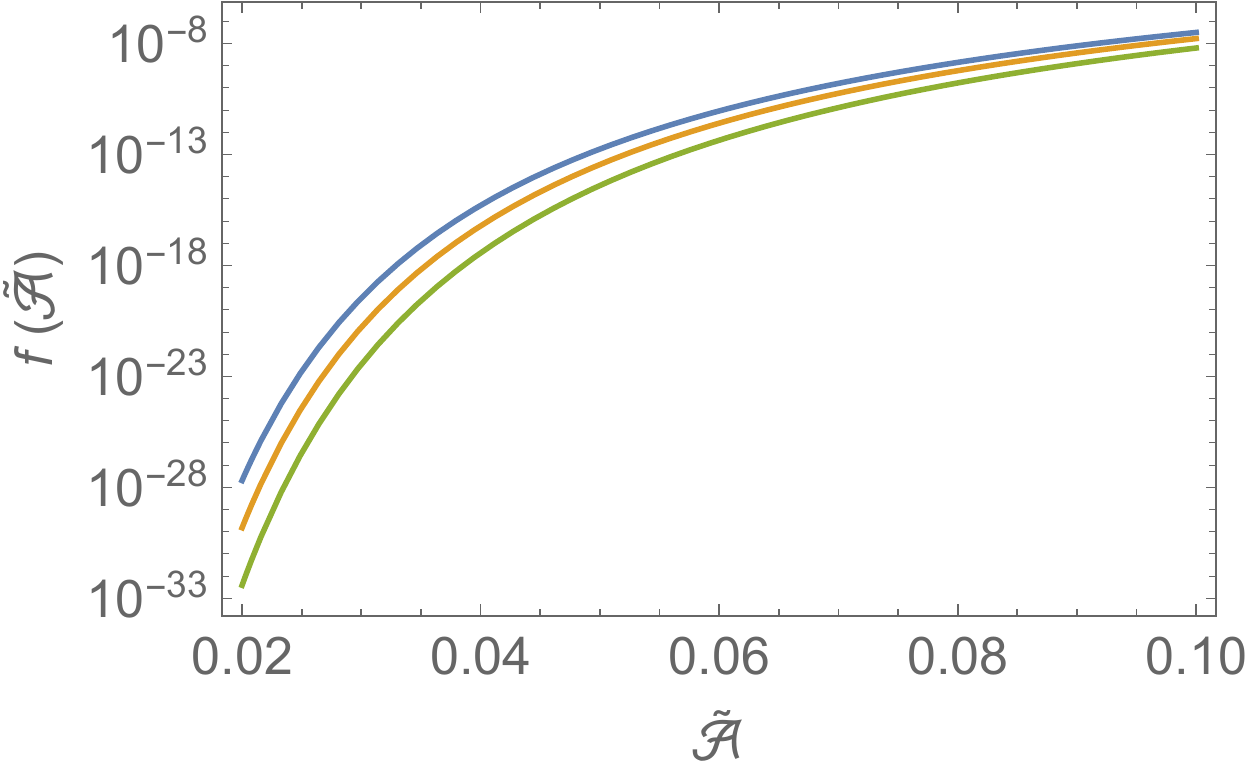}
\end{center}
\caption{
 The function $f(\widetilde{\mathcal A})$. We have taken $\tilde r_{\rm min}=0.1,0.2$ and $0.3$ from top to bottom.
}
\label{fig:fA}
\end{figure}

Now let us estimate more precisely. We take the following approximate configuration for the vector field with the dominant mode $(\ell,j)=(0,1)$~\cite{Baryakhtar:2017ngi},
\begin{align}
	A_i = - \mathcal A(t) e^{-\tilde r} \begin{pmatrix} \cos(\mu_A t) \\ \sin(\mu_A t) \\ 0 \end{pmatrix},~~~
	A_0 = \mathcal A(t) \alpha e^{-\tilde r} \sin(\theta) \cos(\mu_A t-\varphi),
\end{align}
where $\theta$ is the polar angle, $\mathcal A(t)$ denotes the overall amplitude which grows with time during the superradiant phase
and we introduced dimensionless radial coordinate $\tilde r \equiv \alpha\mu_A r$.
This form of the solution is valid except for the near horizon region.
By substituting it, we obtain
\begin{align}
	M_{\rm cloud} =\int d^3 x\left[ \frac{1}{2}(\vec E^2+\vec B^2)+\frac{\mu_A^2}{2} (A_0^2+A_i^2)\right]
	\simeq \pi (\mu_A \alpha)^{-3}\mu_A^2\mathcal A^2,
\end{align}
neglecting terms suppressed by powers of $\alpha$. From this expression we may define the effective volume of the cloud as $\mathcal V \equiv \pi(\mu_A \alpha)^{-3}$.
On the other hand, the Schwinger production rate is given by
\begin{align}
	\int \Gamma_S(E)\,d^3 x \simeq (\mu_A \alpha)^{-3} m_e^4 \widetilde A^2 \, f(\widetilde {\mathcal A}),~~~~~~
	 f(\widetilde {\mathcal A}) \equiv \int_{\tilde r_{\rm min}}^{\infty} d\tilde r \tilde r^2 e^{-2\tilde r}\exp\left(-\frac{e^{\tilde r}}{\widetilde {\mathcal A}} \right),
\end{align}
to the leading order in $\alpha$, where we have defined dimensionless vector amplitude $\widetilde{\mathcal A}\equiv e\mu_A \mathcal A/(\pi m_e^2)$.
Therefore, from (\ref{Mdot_schwinger}), the superradiance rate becomes comparable to the Schwinger production rate when
\begin{align}
	f(\widetilde {\mathcal A}) \simeq \frac{10^{-3} \pi^2 \mu_A}{4\alpha_e m_e},  \label{fA}
\end{align}
where we have used $\omega_I \sim 10^{-3} \mu_A$ for maximal growth rate.
The function $f(\widetilde{\mathcal A})$ is plotted in Fig.~\ref{fig:fA}.
%It is not easy to precisely estimate $\mathcal E_e$, because the kinematics of produced electron-positron pair is complicated due to the external electric and magnetic field, gravity and interactions with themselves or surrounding plasma. 
For the case of the standard model photon around PBHs studied in Sec.~\ref{sec:SM}, we are interested in the range $10^{-8}\,{\rm eV}\lesssim \mu_A \lesssim 10^{-5}\,{\rm eV}$.
In this case Eq.~(\ref{fA}) is satisfied for $\widetilde{\mathcal A}=\widetilde{\mathcal A}_{\rm max} \sim 0.05$.
As far as we are only interested in the value of $\widetilde{\mathcal A}$ at which Eq.~(\ref{fA}) is satisfied, the ambiguity on the energy loss process is not so important since $f(\widetilde{\mathcal A})$ is an exponentially sensitive function of $\widetilde{\mathcal A}$.
It implies that the local energy density of the vector cloud is bounded by
\begin{align}
    \rho_A^{\rm max} \equiv \widetilde{\mathcal A}_{\rm max}^2\left(\frac{\pi m_e^2}{e}\right)^2.  \label{rhoAmax}
\end{align}
The energy extraction rate from the cloud and black hole system is saturated at $\dot M_{\rm tot} \sim - 2\omega_I \rho_{A}^{\rm max} \mathcal V$.
Note that as is seen from Eq.\,\eqref{eq:Euler_Heisenberg}, for $E > \pi m_e^2 / e$, the nonlinear effect in the effective Lagrangian becomes larger and the effective mass may changed by $\mathcal O(1)$.
After all, the effect of the Schwinger pair production constrains the efficiency of the superradiance and the energy density of the vector boson cloud is bounded as (\ref{rhoAmax}), although still there is a gradual energy loss due to the pair production. %Note that the produced electron positron pair is soon accelerated to a relativistic speed by the electric field and escape from the black hole gravity.
In Sec.~\ref{sec:app} we discuss phenomenological implications in the context of the photon superradiance around primordial black holes.

%For a Kerr black hole with angular momentum $J_{\rm BH} \equiv a M_{\rm BH}$, if we ignore the extraction effect, the energy density when all the angular momentum is extracted is roughly
%\begin{align}
%    \rho_s \simeq \frac{a M}{(M / M_\text{Pl}^2)^3} = a (0.1\,\text{GeV})^4 \left(\frac{M_\odot}{M}\right)^2.
%\end{align}
%Thus, depending on the electron mass, $\rho_u < \rho_s$ and the efficient extraction may not occur.

%%%%%%%%%%%%%%%%
%%%%%%%%%%%%%%%%

A few comments are in order.
First, we discuss the validity of the use of the Schwinger pair production rate calculated for a static electric field. If the one electron pair creation rate is much larger than the oscillation frequency of the vector boson, our assumption is justified. The former is\,\footnote{For the particle production by the broad parametric resonance, which is discussed in Sec.\,\ref{sec:pp_scalar}, it is known that the particle creation occurs during a single oscillation of the heavier particle\,\cite{Kofman:1997yn}. It is possible that the same is the case for the Schwinger process. If so, $\Gamma_{e^+e^-} = m_e^{-1}$, which is much larger than Eq.\,\eqref{eq:gamee}. In any case, the constant field approximation is valid.}
\begin{align}
\label{eq:gamee}
    \Gamma_{e^+e^-} \sim \int d^3x \Gamma_S,
\end{align}
and the latter is around $\mu_A$. Therefore, for
\begin{align}
    %\Gamma_{e^+e^-} > \mu_A \iff E > \frac{\mu_A^2}{e},
    \widetilde{\mathcal A}^2 f(\widetilde{\mathcal A}) > \frac{\alpha^2 \mu_A^4}{m_e^4},
\end{align}
the constant field approximation is good enough. Actually this is well satisfied around $\widetilde{\mathcal A} \sim \widetilde{\mathcal A}_{\rm max}$ for the case of photon superradiance around PBHs mentioned above.
%Comparing it with Eq.\,\eqref{eq:Schwinger_rate}, the assumption is consistent when the particle production is efficient if $m_e > \mu_A$.

Second, we comment on the Pauli blocking effect. If the electron and positron in the cloud were confined and abundant, the Schwinger process would be stopped due to the Pauli blocking. However, if the electron-positron abundance is high enough, the pair annihilation process also occurs. This can put the upper bound on the number density of the electron and positron. Let $E_e$ denote a typical energy of the electron/positron and we express the electron/positron number density as,
\begin{align}
    n_{e^+}(E_e) = n_{e^-}(E_e) \equiv c E_e^3.
\end{align}
Then the pair-annihilation rate is
\begin{align}
\Gamma_\text{ann} \sim \int d^3x\,n_{e^+}n_{e^-} \sigma_{\rm ann} v \sim \alpha_e c^2 \frac{m_e^4}{\mu_A^3}
\end{align}
for $E_e \sim m_e$, where $\sigma_{\rm ann}$ is the annihilation cross section and $v$ is the relative velocity. The annihilation rate is smaller than the supply of the electron-positron pair, Eq.\,\eqref{eq:gamee}, if
\begin{align}
    c^2 < \frac{\widetilde{\mathcal A}^2 f(\widetilde{\mathcal{A}})}{\alpha^3\alpha_e}
    \ll 1.
\end{align}
Otherwise, the annihilation rate exceeds the Schwinger production rate. Since this crossover happens at $c\ll 1$, we conclude that the Pauli blocking does not affect our estimation.\footnote{
    Here we have neglected $e^+$ and $e^-$ interaction with the background plasma. It is sufficient, however, since our purpose here is to just show that the Pauli blocking is inefficient. 
} The same is also true even if the electrons/positrons are highly relativistic: $E_e\gg m_e$.

%%%%%%%%%%%%%%%%%%%%%%%%%%%%%%%%%%%%%%

\bibliography{apssamp}% Produces the bibliography via BibTeX.

\end{document}